\documentclass[preprint, 12pt, 5p, twocolumn, sort&compress]{elsarticle}

\usepackage[utf8]{inputenc}
\usepackage{epsfig}
\usepackage{subcaption}
\usepackage{calc}
\usepackage{amssymb}
\usepackage{amstext}
\usepackage{amsmath}
\usepackage{amsthm}
\usepackage{multicol}
\usepackage{pslatex}
\usepackage[bottom]{footmisc}
\usepackage[switch]{lineno}

\makeatletter
\newcommand{\removelatexerror}{\let\@latex@error\@gobble}
\makeatother

\newcommand\sample{\leftarrow\joinrel\smalldollar}

\makeatletter
\newcommand{\smalldollar}{\mathrel{\mathpalette\small@dollar\relax}}
\newcommand{\small@dollar}[2]{%
  \vcenter{\hbox{%
    $#1\textnormal{\fontsize{0.7\dimexpr\f@size pt}{0}\selectfont\$}$%
  }}%
}
\makeatother

\usepackage[ruled,longend,titlenotnumbered,linesnumbered]{algorithm2e}

\NoCaptionOfAlgo

\newtheorem{definition}{Definition}

\journal{Journal of Information Security and Applications}

\begin{document}

\begin{frontmatter}
\title{Cryptanalysis of Cancelable Biometrics Vault}

\author[1]{Patrick Lacharme}

\affiliation[1]{organization={Normandie Univ, UNICAEN, ENSICAEN, CNRS, GREYC},     
            postcode={14000},
            city={Caen},
            country={France}}

\author[2]{Kevin Thiry-Atighehchi}

\affiliation[2]{organization={UCA, CNRS, Mines de Saint-Etienne, Clermont-Auvergne-INP, LIMOS},
            postcode={63000},
            city={Clermont-Ferrand},
            country={France}}

\begin{abstract}
Cancelable Biometrics (CB) stands for a range of biometric transformation 
schemes combining biometrics with user specific tokens to generate secure 
templates. Required properties are the irreversibility, unlikability and 
recognition accuracy of templates while making their revocation possible. 
In biometrics, a key-binding scheme is used for protecting a cryptographic 
key using a biometric data. 
The key can be recomputed only if a correct biometric data is acquired 
during authentication. Applications of key-binding schemes are
typically disk encryption, where the cryptographic key is used to 
encrypt and decrypt the disk.
In this paper, we cryptanalyze a recent key-binding scheme, called 
Cancelable Biometrics Vault (CBV) based on cancelable biometrics. 
More precisely, the introduced cancelable transformation, called BioEncoding 
scheme, for instantiating the CBV framework is attacked in terms of 
reversibility and linkability of templates. Subsequently, our linkability 
attack enables to recover the key in the vault without additional
assumptions. 
Our cryptanalysis introduces a new perspective by uncovering the CBV scheme's revocability and linkability vulnerabilities, which were not previously identified in comparable biometric-based key-binding schemes.
\end{abstract}


\end{frontmatter}





\section{Introduction}
\label{sec:introduction}


Biometric authentication is gaining popularity as a replacement for traditional 
password-based authentication in identity management systems. This is due to 
the uniqueness of biometric data and the added benefit of it being easy to use and
impossible to lose or forget.
Biometric recognition establishes the identity of a person using his
physiological or behavioral modalities.
Physiological modalities like fingerprints, iris, and facial biometrics are predominantly used. Nevertheless, behavioral data based on handwritten signature, voice, or keystroke dynamics, which are also unique to each individual, find usage in certain applications.
Biometric systems can be embedded in many devices, as a smartcard, and are 
deployed for various purposes such as border control, 
authentication for e-banking, or smartphone unlocking, sometimes in combination  
with a second authentication factor.

In such a biometric authentication system, an enrollment biometric
template is acquired from a sensor and a feature extraction process, and 
subsequently stored in a template database. 
Sometimes, several templates are enrolled during this step, in combination with
a quality assessment process. For every authentication attempt, 
a fresh template is acquired and compared to the enrollment template 
using a specific distance metric and a predefined threshold.
If the difference between these two templates falls below the threshold, it is inferred that the fresh template originates from the same individual, thereby validating the claimed identity.
Such a verification system can make two types of errors -- false acceptance and false rejection. These can result from a flawed enrollment process, noise introduced by the sensor, or degraded environmental conditions.
A genuine individual with a template that significantly differs from the enrollment template may be rejected. Conversely, a false acceptance occurs when an impostor is incorrectly identified as genuine by the system.

Another use of biometrics can be found in disk encryption, where a cryptographic key for disk encryption/decryption is derived from the biometric data. Here, the authentication is only implicit -- a newly extracted biometric data enables the retrieval of the decryption key if it closely matches the original data. Key Binding schemes, developed specifically for such applications, will be explained in greater detail below.

Owing to their personal and sensitive nature, biometric data require strong security requirements. 
The unprotected storage of biometric data can result 
in important privacy and security breaches.
Furthermore, unlike passwords or PIN codes, biometric data are irrevocable, meaning they cannot be changed in the event of a security breach. Another challenge in biometric authentication is its vulnerability when systems share the same biometric data (up to a perturbation); if one system is compromised, others become vulnerable as well. Moreover, using the same biometric data across various systems can potentially facilitate user tracking, as cross-matching between databases becomes possible.

Complicating matters, biometric data are the product of sensor measurements, and are 
then subject to noise. This implies that traditional cryptographic mechanisms, such 
as hash functions, are not really appropriate for the protection and use of such data, 
contrary to passwords. Indeed, cryptographic hash functions generate completely 
different digests from two inputs with minor differences. As a consequence, 
fingerprints from fuzzy data cannot be compared.
In addition, symmetric encryption schemes do not protect them during 
the authentication step, because the comparison between biometric templates cannot 
be performed in the encrypted domain.

In light of these challenges, specialized biometric template protection (BTP) 
schemes are designed to secure 
these data, using various schemes as described in some surveys~\cite{jnn08,ru11}. 
These schemes are generally referred to as Biometric Privacy-Enhancing Technologies.
They are based on feature transformations such as cancelable
transformations, or rely on tailored cryptographic approaches, as detailed in
the next section.
A recent survey of Privacy-Enhancing Technologies applied to biometrics is 
proposed in~\cite{mrtvb22}.

Biometric template protection schemes should verify several properties 
such as non-invertibility and unlinkability, while preserving recognition accuracy 
of the system, as outlined in the ISO/IEC Standard
24745 on Biometric Information Protection.
Non-invertibility refers to the difficulty of reconstructing the original biometric data from the protected template. The achievement of this property depends on the particular scheme and the underlying security model, leading to various formalizations.
A general approach considers that given a biometric template, it should be
impossible to recover any information about the original biometric data.
In this paper,
we use notions of preimages, which are more adapted for the description of the
attack. 
Unlinkability  refers to the difficulty of determining whether two protected
templates are derived from two different people or not. 
Finally, the success rate of the authentication should not experience a decline due to the biometric template protection scheme.

Nevertheless, the security of many cancelable transformations is not evaluated within a formal security model and instead relies only on informal arguments. These weaknesses are often exploited for comprehensive cryptanalysis or, at the very least, for the construction of preimages.

Dedicated cryptographic approaches are employed for biometric template protection, for key generation from biometric features, or for securing a key using biometric features. In this latter case, these schemes, referred to as key-binding schemes, typically store public information known as 'helper data'. This helper data is used in conjunction with the fresh template to recover the key.

The security models are founded on various assumptions, and numerous attacks have been proposed against these cryptographic schemes. Detailed cryptanalysis of cancelable transformations and cryptographic schemes for biometric template protection will be discussed in the following section.
Naturally, in the case of key-binding schemes, the best attack consists 
in retrieving the key from the helper data, without the 
 knowledge of the biometric template used for the protection of the key.

Recently, Ouda, Nandakumar and Ross presented a new key-binding scheme, 
called cancelable biometrics vault (CBV) in~\cite{onr20}. This scheme 
is based on the principle of Chaffing and Winnowing, by transforming either
the enrollment template or a fake template, depending on each bit of the 
cryptographic key~$\kappa$. It uses a 
cancelable transformation on a binary biometric template, where each 
generated templates, called biocodes, are used to encode a bit of the key~$\kappa$. 
The algorithm returns a biometric key $k_{bio}$ (the helper data) 
used to recover the original key $\kappa$ with a fresh template from the 
same user. Experiments are realized with iris templates from the CASIA 
database and provide good performance.  

However, the security of this scheme is only analyzed at high level and
no formalization or security proofs are presented by the authors. 
In this paper, we propose a cryptanalysis of this key-binding scheme by 
recovering the key $\kappa$ from the helper data  $k_{bio}$. 
The attack is based on the linkability of templates generated from the
same people, which reduces drastically the size of the key space.
This approach of attack was not previously proposed
in comparable biometric-based key-binding schemes.

The paper is organized as follows. Section~\ref{sec:relatedworks} presents a survey of biometric template protection schemes, along with an exploration of the main attacks on them. In Section~\ref{sec:preliminaries}, we provide a description of the Cancelable Biometrics Vault (CBV) scheme. Section~\ref{sec:Key-Recovery Attack on CBV} details the proposed attack and includes numerical examples for clarity and a security discussion on the scheme.


\section{Related works}
\label{sec:relatedworks}


In this section we describe the two main approaches for biometric template
protection, which are the cancelable transformations and the cryptographic
approaches. At high level, these approaches seem totally independent, but
cryptographic approaches generally require binary templates and cancelable
transformations can sometimes be used as a binary process before the
cryptographic scheme.


\subsection{Cancelable transformations}
\label{ref:subsec:cancelable}

Cancelable transformations have been proposed in~\cite{rcb01} for biometric data
protection. Such a transformation is applied on the biometric data, with
an additional data, called seed or token. During the authentication step,
the transformation is applied on the fresh template with the same token 
and the distance between the two transformed templates is compared to a threshold 
to decide the validity of the claimed identity. 
The token can be secret or public, but the security analysis of a cancelable 
transformation generally considers that the token is known by an attacker 
(stolen token scenario).

Many transformations have been proposed, as the biohashing 
algorithm~\cite{tng04,tgn06,tkl08} and, more generally, schemes using  
random projection~\cite{fyj10,ppcr11,wp10}. 
These transformations use the product of the feature vector, derived from
the biometric template, by a pseudo random matrix derived from the
token and the obtained vector is generally binarized during a last step.
The constrution is based on the Johnson and Lindenstrauss Lemma~\cite{jl84},
nevertheless, these transformations are vulnerable to several attacks,
as linear programming approaches~\cite{nnj10,fly14,tkae16}.
These attacks modelize the original template as a solution of a set of 
linear inequalities, which is generally easy to solve.  

An other approach to attack these transformations uses genetic
algorithms to find preimages, as proposed and implemented on 
fingerprints~\cite{lcr13} and iris~\cite{grgfo13}. 
Genetic algorithms have been later applied on
other cancelable transformations as presented in~\cite{rgb15,djt19}.

Bloom filters have also been investigated for the construction of
cancelable transformation, first on iris biometrics~\cite{rbbb14} and later on face 
biometrics~\cite{rgrfb14}. Nevertheless, these bloom filters based 
transformation are known to not ensure unlinkability property~\cite{bmr15}.

Transformations based on fingerprint minutiae positions are proposed by
\cite{rccb07}, but they are vulnerable to attacks on non-invertibility, 
via record multiplicity, as presented in~\cite{fscz08}.
More recent transformations use stronger primitive as locality sensitive
hashing function, with index-of-max~\cite{c02,jhlkt18}, but they have also
been cryptanalyzed with linear and geometric programming methods~\cite{gkla20}.

\subsection{Cryptographic approaches}
\label{subsec:cryptographicapproaches}

Fuzzy extractors~\cite{drs04}, are intended to generate a 
cryptographic key from a fuzzy data, generalizing older constructions
as fuzzy commitments~\cite{jw99} and fuzzy vaults~\cite{js02}, based
of error-correcting codes theory, used to recover the key from
a fresh data and a non-sensitive helper data.
Key-binding schemes, where a biometric data is binding with a cryptographic 
key such that neither the key nor the biometric data can be recovered,
can also be designed with the previous primitive, under the
same security model. The fuzzy commitment scheme have been first implemented 
on iriscodes, using a combination of Reed-Solomon codes and Hadamard 
codes~\cite{had06} or using BCH codes~\cite{yv07}. A theorical analysis
on fuzzy commitment schemes is proposed in~\cite{bcckz08}. 
The first implementation of fuzzy vault uses minutiae from fingerprints,
as proposed in~\cite{njp07}. 

The security of these schemes is variable. Thus, it was generally considered 
that fuzzy commitment is secure, if the biometric templates are uniformly and 
independently distributed. Nevertheless this scheme leaks information on 
both secret and biometric template if this assumption is not verified 
\cite{iw10,zkvb11}. Moreover, the security is not ensured with multiple
commitments, which can been combined to reconstruct secrets~\cite{b04}
or provide privacy leakages. If the result of~\cite{b04} was essentially 
theorical, several attacks have been proposed later, as for example the
undistinguishability attack presented in~\cite{stp09} which exploits that in 
a linear code, the sum of two codewords is also a codeword of the code.
An improved version, based on a public permutation is proposed in
~\cite{kbkbv11}, but this variant has been also cryptanalyzed in~\cite{t14}.
Similar attacks on multiple commitments are also known on the fuzzy vault scheme,
as presented in~\cite{stp09} and implemented
on fingerprints in~\cite{kbkbv11} and later in~\cite{ba13},
giving rise to improved versions~\cite{tmm15,rtwb16}. 

A new generation of fuzzy extractors has been recently proposed. Thus, a 
fuzzy extractor, based on the ancient fuzzy commitment construction of~\cite{jw99},
but with a security based on the hardness of the Learning with Errors problem 
is proposed in~\cite{fmr13}. This scheme was proven non reusable~\cite{acek17},
but authors provide some changes to obtain weakly reusable and strongly
reusable schemes. Other reusable fuzzy extractors schemes have been presented, 
as in~\cite{wl18,abccfgs18,cfprs21}, with security models based on the hardness
of various problems, and are secure against computationally bounded adversaries
if the entropy of biometric data is sufficient. Thus, in~\cite{wl18}, a
Symmetric Key Encapsulate Mechanism is introduced, constructed under 
the DDH assumption security model. Most of these schemes have 
not been implemented on real biometric data, excepted for 
schemes  of~\cite{cfprs21}, based on the sample-then-lock approach using 
hash functions or HMAC, for which implementation are proposed 
on Iris biometrics in~\cite{ssf19} and face biometrics in~\cite{uyckl21}.

In conclusion for this section, most biometric template protection schemes, whether they are cancelable transformations or cryptographic approaches, are considered insecure or not suited to real biometric data. However, the new generation of fuzzy extractors appears to offer a solution tailored to the challenges of biometric data protection.


\section{Preliminaries}
\label{sec:preliminaries}


Before presenting the cancelable biometrics vault (CBV) in detail, in 
the subsection~\ref{subsec:cbv}, the first 
subsection gives the generic definitions of both cancelable biometric 
transformation and key-binding schemes. Finally, different attacks,
as reversibility or linkability are described in the subsection~\ref{subsec:atttackmodels}.


\subsection{Biometric Transformations and Key-Binding}
\label{subsec:keybinfingdefinition}


In the following, we let $(\mathcal{BT},D_H)$ and $(\mathcal{BC},D_H)$ be 
two metric spaces, where $\mathcal{BT}$ and $\mathcal{BC}$ represent the 
template space and biocode space, respectively, and $D_H$ is the Hamming 
distance defined by 
$$D_H(x, y) = |\{i\ |\ x_i \neq y_i\}|.$$ 
A cancelable biometric scheme is used for the protection of a biometric
data. It is used during the enrollment and during the authentication
step because biometric data are compared in the  biocode space
$(\mathcal{BC},D_H)$.
In the following, all parameters are public because attack models
on these transformations generally consider no secret and because there 
are no secret key in transformations used in the CBV scheme.

\begin{definition}
Let $\mathcal{P}$ be the paramaters space, representing the set of tokens to 
be assigned to users. A cancelable biometric scheme is a tuple of 
deterministic polynomial time algorithms $\Pi:=(\mathcal{T}, \mathcal{V})$, 
where
\begin{itemize}
\item $\mathcal{T}$ is the transformation of the system, that takes an 
enrollment biometric template $x^e \in \mathcal{BT}$ and a set of public 
parameter $pp \in \mathcal{P}$ 
as input, and returns a biocode $b=\mathcal{T}(pp,x^e) \in \mathcal{BC}$.
\item $\mathcal{V}$ is the verifier of the system, that takes a 
biometric fresh template $x^a \in \mathcal{BT}$, a biocode
$b$ = $\mathcal{T}(pp,x^e)$ and a threshold $\tau$ as input 
and returns $True$ if $D_H(b, \mathcal{T}(pp,x^a)) \leq \tau $, and
returns $False$ otherwise, where $D_H$ is an adapted distance.
\end{itemize} 
\end{definition}

The following definition of the key-binding scheme is presented in the 
general case, without cancelable transformations. 
A key-binding scheme combines a cryptographic key $\kappa$ and a biometric 
data into an helper data, called here cryptographic key $\kappa_{bio}$.
From $\kappa_{bio}$ and a biometric data close to the original, $\kappa$
can be recovered. These schemes have been formalized in \cite{drs04}.

\begin{definition}
Let $\mathcal{K}$ the space of keys. A key-binding scheme is a tuple of 
deterministic polynomial time algorithms $\Xi:=(\mathcal{KB}, \mathcal{KR})$, 
where
\begin{itemize}
\item $\mathcal{KB}$ is the key-binding algorithm that takes a cryptographic 
key $\kappa \in \mathcal{K}$, an enrollment biometric template 
$x^e\in \mathcal{BT}$ and a set of public parameters $pp\in \mathcal{P}$ 
as input, and returns a biometric key $k_{bio}=\mathcal{KB}(pp,x^e,\kappa)$.
\item $\mathcal{KR}$ is the key-release algorithm, that takes a 
biometric fresh template $x^a \in \mathcal{BT}$, a biometric key 
$k_{bio}=\mathcal{KB}(pp,x^e,\kappa)$, a threshold $\tau$ and a set of public 
parameters $pp\in \mathcal{P}$ as input and returns $\kappa$ if 
$D_H(x^e, x^a) \leq \tau $, and returns $False$ otherwise, where $D_H$ is an 
adapted distance.
\end{itemize} 
\end{definition}

The minimum security level for a key-binding scheme requires that it must not 
be possible to retrieve the key $\kappa$ or the biometric data $x^e$ 
from the biometric key $k_{bio}$. 
It should be also secure in case of reusability, which considers multiple
enrollments of the people.


\subsection{The Cancelable Biometrics Vault Scheme}
\label{subsec:cbv}


The CBV scheme has been proposed in~\cite{onr20}. The following two subsections 
give a description of its ingredients  and use the same notations
of the original paper.
For example, we use the term \emph{biometric key} for $k_{bio}$, instead of the 
term \emph{helper data}.
At high level, Key-Binding in CBV is realized in two steps:
\begin{enumerate}
\item[($i$)] A number $n$ of public parameter sets are defined for deriving a 
family of $n$ CB schemes; 
\item[($ii$)] Each instantiated CB scheme then serves for locking one bit of 
the key.\\
\end{enumerate}

\subsubsection{The Key-Binding Scheme in CBV}
\label{subsec:keybinding}

\begin{figure*}[h]
\begin{center}
\includegraphics[width=9cm]{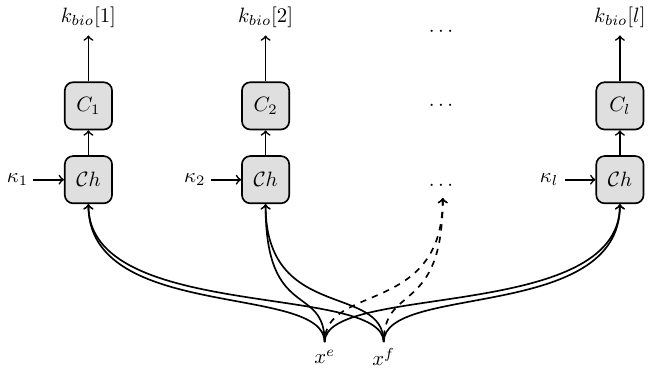}
\end{center}
\caption{The chaffing process of the $\mathcal{KB}$ operation, where $\mathcal{C}h$ denotes the Chaff selector.
Depending on the value of the selector bit $\kappa_i$, $\mathcal{C}h$ returns either the genuine template $x^e$ (if the selector bit is 1) or the fake template $x^f$ (if the selector bit is 0).
}
\label{chaff_fig}
\end{figure*}

The key-binding algorithm $\mathcal{KB}$ uses the bit $\kappa_i$ of the
$l$-bits key $\kappa$ as a decision system to put the $i$-th cancelable transformation
of the enrollment template $x^e$, noted by $C_i(x^e)$ into $k_{bio}[i]$
or the the $i$-th cancelable transformation of a fake template $x^f$, 
noted by $C_i(x^f)$ into $k_{bio}[i]$. This overall transformation is referred to as the Chaffing process and is depicted Figure~\ref{chaff_fig}. 
The result of this operation is the locked key $k_{bio}$ which is stored along with the $l$ cancelable transformations and the hash value of $\kappa$. 
In the following, the set of $l$
cancelable transformation will be transformed into a randomized cancelable 
transformation. The hash function is a cryptographic hash function as
SHA-256.

In the original algorithm, it is proposed to generate $x^f$ at random
or by a permutation of reals components of $x^e$. Without loss of 
generalities, we assume in the rest of the paper that it is generated at random.
Moreover, given an $l$-bit key 
$\kappa$, the size of the corresponding biometric key $k_{bio}$ is $l$ times the output size of a cancelable transformation $C_i$.\\

\begingroup
\removelatexerror
\begin{algorithm}[H]
\caption{\bf The key-binding algorithm $\mathcal{KB}$}
\KwIn{the enrollment template $x^e$, a fake template $x^f$, a cryptographic 
key $\kappa$ of $l$ bits and a set of $l$ cancelable transforms $C_i$, with 
$i$ between $1$ and $l$ and a cryptographic hash function $H$.}
\KwOut{the biometric key $k_{bio}$ and $H(\kappa)$.}
\ForAll{i between $1$ to $l$}
{
\uIf{$\kappa_i$ = 1}{$k_{bio}[i]\gets C_i(x^e)$}
\Else{$k_{bio}[i]\gets C_i(x^f)$}
}
\Return $k_{bio}$\\
\end{algorithm}
\endgroup

~\\

\begin{figure*}
\begin{center}
\includegraphics[width=9cm]{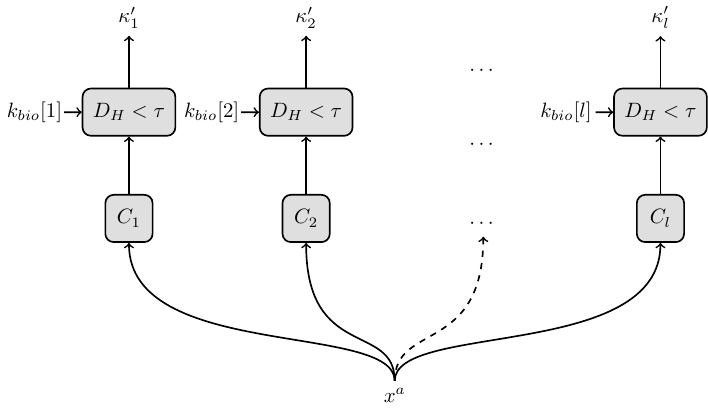}
\end{center}
\caption{
The winnowing process in the $\mathcal{KR}$ operation involves distinct transformations of the fresh template, which separate the genuine transformations (represented by a bit 1 in the unlocked key) from the fake transformations (represented by a bit 0 in the unlocked key).
}
\label{winnowing_fig}
\end{figure*}

The key-release algorithm $\mathcal{KR}$ computes the distance between 
each word $k_{bio}[i]$ with the cancelable transformation $C_i(x^a)$ of the fresh template 
$x^a$. It considers that if the distance is low, then $k_{bio}[i]$
is $C_i(x^e)$ and $\kappa'_i = 1$, and if the distance is high, then $k_{bio}[i]$
is $C_i(x^f)$ and $\kappa'_i = 0$. 
This transformation is referred to as the Winnowing process and is depicted Figure~\ref{winnowing_fig}.
At the end of this algorithm, the key 
$\kappa'$ is checked against the hash value $H(\kappa)$, and if the hash values
are equal, $\kappa$ has been correctly recovered.\\

\begingroup
\removelatexerror
\begin{algorithm}[H]
\caption{\bf The key-release algorithm  $\mathcal{KR}$}
\KwIn{the biometric key $k_{bio}$, the fresh template $x^a$, $h=H(\kappa)$ the hashed key,
a threshold $\tau$ and the same set of $l$ cancelable transforms $C_i$, 
with $i$ between $1$ and $l$ used in the key-binding scheme.}
\KwOut{The key $\kappa$ or Failure.}
\For{$i=1$ \KwTo $l$}
{
Compute $C_i(x^a)$
}
\For{$i=1$ \KwTo $l$}
{
\uIf{$D_H(C_i(x^a), k_{bio}[i]) \geq \tau$} {$\kappa'_i\gets 0$}
\Else{$\kappa'_i\gets 1$}
}
\uIf{$h = H(\kappa')$}{\Return $\kappa'$}
\Else{\Return Failure}
\end{algorithm}
\endgroup

~\\

This key-release algorithm is very restrictive : given the enrollment
template $x^e$ and the fresh template $x^a$, if there is only
one $i$ between $1$ to $l$ such that $\kappa_i = 1$ and 
$D_H(C_i(x^e), C_i(x^a))\geq\tau$, then the algorithm fails. It requires a
strong intraclass stability in biometric templates after transformation.

The security of this key-binding scheme is directly related to the security
of the cancelable transformation, which must be non invertible
and unlinkable.\\

\subsubsection{The Cancelable Transformation in CBV}
\label{subsec:cancleabletransformcbv}

A cancelable transform has two algorithms : an enrollment algorithm and
a authentication algorithm.

During the enrollment, the objective of the algorithm is the 
creation of a biocode from a biometric template $x^e$. 
This biocode is used in the authentication procedure. A basic security 
requirement is that the template $x^e$ cannot be recovered from the biocode.

\begin{figure*}
\begin{center}
\includegraphics[width=15cm]{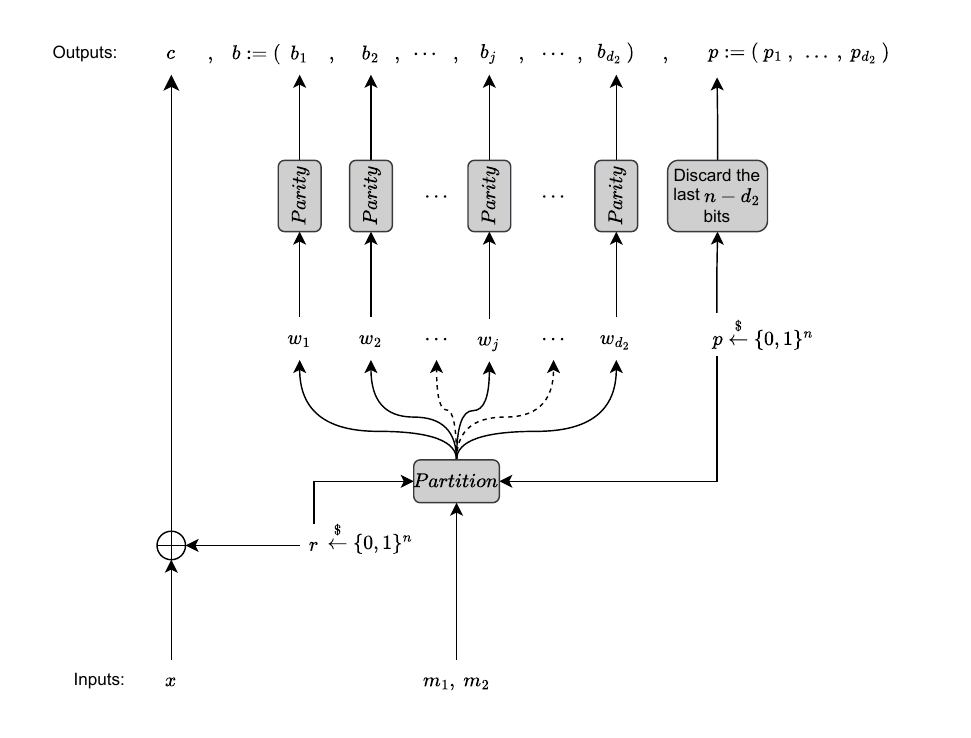}
\end{center}
\caption{Cancelable Transformation of CBV}
\label{fig:transf}
\end{figure*}

The enrollment algorithm generate at random two $n$-bits integers $r$ and $p$.
The integer $r$ is divided in words of $m_1$ bits or $m_2$ bits,
depending to the bits of $p$. Then, the bits of each words $w_j$ of $r$
are xored into a bit $b_j$ and the template $x^e$ is xored with $r$
into $c$. The result of this enrollment is a set of three integers
$\{b, c, p\}$ ($r$ is not disclosed). An illustration of the transformation is depicted Figure~\ref{fig:transf}.

The authentication algorithm attempts to recover $r$ by a xor 
between $x^a$ and $c$. The result $r'$ of this xor is divided in
words of $m_1$ bits or $m_2$ bits, depending to the bits of $p$,
as in the enrollment process. Then, the bits of each words $w'_j$ of $r'$
are xored into a bit $b'_j$ and the authentication is successful if
$b$ and $b'$ are close, using the Hamming Distance. These algorithms
are detailed below.\\
 
\begingroup
\removelatexerror
\begin{algorithm}[H]
\caption{\textbf{The cancelable transform (enrollment)}}
\KwIn{an enrollment template $x^e$ of length $n$, two word-lengths $m_1$ 
and $m_2$.}
\KwOut{a biocode $\{b, c, p\}$.}
$r, p\sample \{0,1\}^n$\\
Divide $r$ into $m$-bits words, where $m$ = $m_1$ if $p_i$ = $0$ and $m$ 
= $m_2$ if $p_i$ = $1$ and $m\leq m_2$ for the last word\\
Let $d_2$ be the number of words of $r$. Delete the $n-d_2$ most 
significant bits of $p$ (the length of $p$ is now $d_2$ bits)\\
\ForAll{words $w_j$ in $r$}
{
        $b_j \gets  w_{j,1}\oplus w_{j,2}\oplus\ldots
\oplus w_{j,m}$
}
$c\gets x\oplus r$\\
\Return $b, c, p$\\
\end{algorithm}
\endgroup

Remark : the integer $d_2$ is the number of words in the partition of $r$. It cannot be hard-fixed since it corresponds to the number of steps in a random walk.
According to the Wald's equation~\cite{wald1944cumulative}, the averaged $d_2$ is actually greater than or equal to $\frac{2n}{m_1+m_2}$.\\

\begingroup
\removelatexerror
\begin{algorithm}[H]
\caption{\textbf{The cancelable transform (authentication)}}
\KwIn{a fresh template $x^a$, a biocode $\{b, c, p\}$ and a threshold $\tau$.}
\KwOut{Success or Failure.}
$r'\gets x^a\oplus c$\\
Divide $r'$ into $d_2$ $m$-bits words, where $m$ = $m_1$ if 
$p_i$ = $0$ and $m$ = $m_2$ if $p_i$ = $1$ and $m\leq m_2$ for the last word\\
\ForAll{words $w'$ in $r'$}
{
        $b'_j \gets  w'_{j,1}\oplus w'_{j,2}\oplus\ldots
\oplus w'_{j,m}$
}
\uIf{$D_H(b, b')\leq \tau$}{\Return Success}
\Else{\Return Failure}
\end{algorithm}
\endgroup

Remark : A biocode $\{b^i, c^i, p^i\}$ corresponds to the cancelable transform $C_i(x)$.
Thus, the biometric key $\kappa_{bio}$ is a set of $l$ biocodes $\{b^i, c^i, p^i\}$ for
$i$ between $1$ and $l$, computed as described above.

\subsection{Attack Models}
\label{subsec:atttackmodels}

Let $\mathcal{U}$ be the set of users of the biometric system, identified by their 
biometric characteristics. A biometric characteristic $u \in \mathcal{U}$ is 
captured and transformed into a biometric template $x$ = $\mathcal{BT}(u)$.
 Due to the inherent noise in the capture, two computations 
 $x=\mathcal{BT}(u)$ and $x' = \mathcal{BT}(u)$ can be different for
 a same user $u$. Conversely, we can have $\mathcal{BT}(u)=\mathcal{BT}(u')$ 
 for two different 
 users $u$ and $u'$ of $\mathcal{U}$. 
Definitions of this section are based on \cite{gkla20}.  

\subsubsection{Reversibility Attacks on CB Schemes}

Let $b=\Pi.\mathcal{T}(pp,x)\in\mathcal{BC}$ be the biocode generated from a template $x\in \mathcal{BT}$ and the public parameter set $pp \in \mathcal{P}$. In an authentication attack, an adversary is given $b$, $pp$, and a threshold value $\tau$, and the adversary tries to find a template $x^*\in \mathcal{BT}$ such that 
$b^*=\Pi.\mathcal{T}(pp,x^*)$ is exactly equal to $b$, or $b^*$ is close to $b$ with respect
to the distance function over $\mathcal{BC}$ and the threshold value $\tau$.
In this case, we say that $x^*$ is a $\tau$-nearby-biocode preimage 
of the biocode $b$.
More formally, we have the following definition. 

\begin{definition}\label{def:preimage}
Let $x\in \mathcal{BT}$ be a template, $\tau$ be a threshold 
and $b=\Pi.\mathcal{T}(pp,x)\in\mathcal{BC}$ be a biocode for some public parameter $pp$. A {\it nearby-biocode preimage} of $b$ with respect to $pp$ and $\tau$ is a template $x^*$ such that
$b^*=\Pi.\mathcal{T}(pp,x^*)$ and
$\Pi.\mathcal{V}(b,b^*,\tau) = True$.
\end{definition}

An adversary $\mathcal{A}$ in an authentication attack is an algorithm that 
takes $pp$ and $b$ = $\Pi.\mathcal{T}(pp,x)$ as input, and that outputs $x^* = \mathcal{A}(pp, b)\in\mathcal{BT}$. In this case the attack is considered as successful if $x^*$ is a nearby-biocode preimage of $b$. Note that this attack, as an authentication attack, do not care about the proximity between $x$ and the constructed preimage $x^*$.

More formally, we have the following definition of an adversary, where $x \sample \mathcal{S}$ 
indicates that $x$ is chosen from the set $\mathcal{S}$ uniformly at random.

\begin{definition}
\textit{
Let $\Pi$ be a cancelable transformation scheme and  $\mathcal{A}$ an adversary for finding a nearby-biocode preimage.
 The success rate of $\mathcal{A}$, denoted by $\mathrm{Rate}^{Auth}_{\mathcal{A}}$, is defined as:
\[
\Pr\left[ \Pi.\mathcal{V}(b,b^*,\tau) = True \;\middle|\;
  \begin{tabular}{@{}l@{}}
   $u \sample \mathcal{U};$\\
   $x \gets \mathcal{BT}(u);$\\
   $pp \gets \mathcal{P}$;\\
   $b \gets \Pi.\mathcal{T}(pp,x);$\\ 
   $x^* \gets \mathcal{A}(pp, b);$\\
   $b^* \gets \Pi.\mathcal{T}(pp,x^*);$
   \end{tabular}
  \right].
\]
}
\end{definition}

An adversary can follow a {\it naive strategy}.
and expected to succeed with probability $\mathrm{FMR}(\tau)$, which is the false accept rate of the system. 
This strategy is also commonly known as the false match rate attack in the literature. A weakness of the scheme, with respect to the false authentication notion, would require better attack strategies, and this motivates the following definition.

\begin{definition}
\textit{
The transformation scheme $\Pi$ is said to have false authentication with advantage $\mathrm{Adv}^{Auth}(\mathcal{A})$ property, if there exists an adversary $\mathcal{A}$ such that $|\mathrm{Rate}^{Auth}_{\mathcal{A}} -  \mathrm{FMR}(\tau)|\ge \mathrm{Adv}^{Auth}(\mathcal{A})$.
If $\mathrm{Adv}^{Auth}(\mathcal{A})$ is negligible for all $\mathcal{A}$, then we say that $\Pi$ does not have false authentication property under the stolen token scenario.}
\end{definition}


\subsubsection{Linkability Attacks on CB Schemes}
Let $b=\Pi.\mathcal{T}(pp,x)\in\mathcal{BC}$ and 
${b^\prime}=\Pi.\mathcal{T}({pp^\prime},{x^\prime})\in\mathcal{BC}$
be two biocodes generated from two templates of $x$ and ${x^\prime} \in \mathcal{BT}$, with the public parameters set $pp$
and ${pp^\prime}$.
In a linkability attack, an adversary tries to find out whether $b$ and ${b^\prime}$ are 
derived from the same user, given $pp$, ${pp^\prime}$, $b$, and ${b^\prime}$. 
An adversary $\mathcal{A}$ in a linkability attack is modelled as an algorithm that 
takes $pp$, ${pp^\prime}$, $b$, and ${b^\prime}$ as input, and that outputs $0$ or $1$,
where the output $1$ indicates that 
the templates $b$ and ${b^\prime}$ are extracted from the same user, and
the output $0$ indicates that 
the templates $b$ and ${b^\prime}$ are extracted from two different users.
We say that the adversary $\mathcal{A}$ is successful, if his conclusion on whether the biocodes are extracted from the same user
is indeed correct.

\begin{definition}
\textit{
Let $\Pi$ be a cancelable transformation scheme and  $\mathcal{A}$ an adversary for a linkability attack.
The success rate of $\mathcal{A}$, denoted by $\mathrm{Rate}^{Link}_{\mathcal{A}}$, is defined as:
\[
\Pr\left[ c' = c \;\middle|\;
  \begin{tabular}{@{}l@{}}
   $u \sample \mathcal{U};$ $x \gets \mathcal{BT}(u);$\\
   $pp \sample \mathcal{P};$ $pp' \sample \mathcal{P};$\\
   $c \sample \{0,1\}$;\\
   $u' \gets u$  if $c=0$;\\
   $u' \sample \mathcal{U}\setminus u$ if $c=1$;\\
   $x' \gets \mathcal{BT}(u');$\\
   $b \gets \Pi.\mathcal{T}(pp,x); {b^\prime} \gets \Pi.\mathcal{T}({pp^\prime},{x^\prime});$\\ 
   $c' \gets \mathcal{A}(pp, b, {pp^\prime}, {b^\prime});$
   \end{tabular}
  \right].
\]
}
\end{definition}

Remark: an adversary can follow a naive strategy by merely sampling a value from $\{0,1\}$.
Under this strategy, the adversary would be expected to succeed with probability $1/2$. 



\subsubsection{Key-Recovery Attacks on KB schemes}

The challenger generates a public parameter set $pp$, chooses a random $\kappa$ from $\mathcal{K}$, a random biometric template $x$ from $\mathcal{BT}$, computes a biometric key $k_{bio} = \Xi.\mathcal{KB}(pp,x,\kappa)$ and sends $k_{bio}$ and $pp$ to the attacker $\mathcal{A}$. The attacker $\mathcal{A}$ then sends $\hat{\kappa}$ back to the challenger.
We say that the adversary $\mathcal{A}$ is successful if $\hat{\kappa} = \kappa$. More formally, we have the following definition.

\begin{definition}
Let $\Xi:=(\mathcal{KB}, \mathcal{KR})$ be a key-binding scheme and  $\mathcal{A}$ an adversary for a key-recovery attack.
The success rate of $\mathcal{A}$, denoted by $\mathrm{Rate}^{Rec}_{\mathcal{A}}$, is defined as: 
\[
\Pr\left[ \hat{\kappa} = \kappa \;\middle|\;
  \begin{tabular}{@{}l@{}}
   $u \sample \mathcal{U};$ $x \gets \mathcal{BT}(u);$\\
   $k_{bio} \gets \Xi.\mathcal{KB}(pp,x,\kappa);$\\ 
   $\hat{\kappa} \gets \mathcal{A}(pp,k_{bio});$
   \end{tabular}
  \right].
\]
\end{definition}

Note that an adversary can follow a {\it first naive strategy} which consists in randomly generating  a key $\kappa^*$ from $\mathcal{K}$ and proposing $\kappa^*$ to the challenger. Under this strategy, the adversary would be expected to succeed with rate $1/\|\mathcal{K}\|$. However, the entropy of a biometric template is known to be much lower than that of a cryptographic key. Hence, a better strategy, the {\it second naive strategy}, is to repeat sampling a biometric template $x^*$ from $\mathcal{BT}$, and computing $\Xi.\mathcal{KR}(pp,x^*, k_{bio})$ until the result is different from False. With this new strategy,
the success rate of $\mathcal{A}$ is~1 with an expected number of $1/\mathrm{FMR}(\tau)$ calls to the $\mathcal{KR}$ algorithm, where $\mathrm{FMR}(\tau)$ is the false accept rate of the system. 

Hence, considering the adversary has no restriction for the access to the algorithms of $\Xi$, an efficient adversary should be able to recover the key in less than $1/\mathrm{FMR}(\tau)$ calls to the $\mathcal{KR}$ algorithm.


 \section{Key-Recovery Attack on CBV}
 \label{sec:Key-Recovery Attack on CBV}


\subsection{Preimage Forgeries}
\label{subsec:Preimage Forgeries}

Given a biocode $\{b, c, p\}$, an attacker knows the overall parity of each 
word $w_j$ of $r$ (for $j$ between $1$ and $d_2$). From this and $c$, the 
attacker deduces the overall parity of each corresponding word of the
template $x^e$. A preimage $x$ with words possessing the same parity is
authenticated because $r' = x\oplus c$, divided into $d_2$ words provides 
$b'=b$.

We observe that $d_2$ bits can serve as adjustment variables for yielding the target word parities. Hence, there are exactly $2^{n-d_2}$ preimages for which $D_H(b, b')$ = $0$
in the authentication procedure and these preimages are easily known by an
attacker having the knowledge of the biocode $\{b, c, p\}$. On the other hand,
the attacker cannot retrieve the original template, used in enrollment, 
among these preimages.

As regards the nearby-biocode preimages (Definition \ref{def:preimage}), up 
to $\tau$ errors can be 
tolerated on $b$. That means that for up to $\tau$ words in $x$, we do not 
care about what may be their parities, hence rising the number of nearby-biocode 
preimages up to $\sum_{i=0}^{\tau} {d_2 \choose i}2^{n-(d_2-i)}$.


\subsection{Linkability Attack}
\label{subsec:Linkability Attack}

In \cite{onr20}, the unlinkability of the transformation is 
only considered with $m_1$ = $m_2$. Nevertheless, the transformation is
also linkable with $m_1\neq m_2$ as detailed in this subsection.

In a first part, we assume that if biocodes come from the same people,
then the original templates are equal and if biocodes come from two
different people, then there is a probability of $0.5$ that bits of
original templates are equal.
This assumption is only used to describe the linkability attack of this subsection and
is not mandatory for the whole attack. In a general case, a distinguisher can only
decide if two biocodes are generated from two templates, which are close or not. 

Let $\{b^1, c^1, p^1\}$ and $\{b^2, c^2, p^2\}$  be two biocodes.
We consider the following distinguisher between these
two biocode. Let $p_1^1$ and $p_1^2$ be the first bit of $p^1$ and $p^2$.
If $p_1^1\neq p_1^2$, the attacker cannot decide if these biocodes
are from the same people or not, with a probability greater than $0.5$. 
In this case, the distinguisher returns Failure. 
If $p_1^1$ = $p_1^2$, then the attacker
computes the parity of the first word of the original template. 
If the parity is different, the attacker can conclude 
that biocodes come from different people.
In the other case, with $p_1^1$ = $p_1^2$ and an equal parity, the 
attacker cannot conclude if biocodes come 
from the same people or from different people with a probability greater 
than $0.5$. In this case, the distinguisher considers the second bits 
$p_2^1$ and $p_2^2$ of $p^1$ and $p^2$ and applies the same strategy 
than previously. Thus, two biocodes generated from two different
peoples are linkable with a probability $\sum_{i=1}^{d_2}1/2^i$.

\begin{algorithm}
\caption{\textbf{Linkability attack on two biocodes}}
\KwIn{two biocodes $\{b^1, c^1, p^1\}$ and $\{b^2, c^2, p^2\}$}
\KwOut{A boolean (1 for same user and 0 for different users) or Failure}
\For{$i=1$ \KwTo $d_2$}{
        \If{$p_i^1\neq p_i^2$}{\Return Failure\\}
        Let $m$ = $m_1$ or $m_2$ depending on the value of $p_i^1 = p_i^2$\\
        \If{$c^1_{i,1}\oplus \ldots \oplus c^1_{i,m}\oplus b^1_i\neq c^2_{i,1}\oplus \ldots \oplus c^1_{2,m}\oplus b^2_i$}{\Return 0\\}
    }
    \Return 1\\
\end{algorithm}

This linkability attack is not directly used in the cryptanalysis of the CBV scheme. 
It is only presented because there is a link with the cryptanalysis of
the modified CBV scheme.


\subsection{Cryptanalysis of the CBV scheme}

In this section, we consider a set $\kappa_{bio}$ of $l$ biocodes 
$\{b^1, c^1, p^1\}, \ldots ,\{b^l, c^l, p^l\}$, generated by a genuine
template $x^e$ and a cryptographic key $\kappa$ with the previous
key-binding scheme. The attacker attempts to distinguish if these biocodes 
have been computed from the same people (genuine template) or not (the 
fake template) and recovers $\kappa$ from $\kappa_{bio}$, without 
additional assumptions. Remark : $l$ 
is the binary size of the cryptographic key $\kappa$, for example $l$ = $128$.  

In a first time, the attacker separates the $l$ biocodes into two groups 
$G_0$ and $G_1$, 
depending on the value of the bit $p_1^i$, for $i$ between $1$ and $l$. 
Consequently, the first group $G_0$ is composed of
biocodes where the length of the first word of $r$ is $m_1$ 
(with $p_1^i = 0$) 
and the second group $G_1$ is composed of biocodes where the length of the first 
word of $r$ is $m_2$ (with $p_1^i = 1$) . 

\begin{figure*}
\begin{center}
\includegraphics[width=19cm]{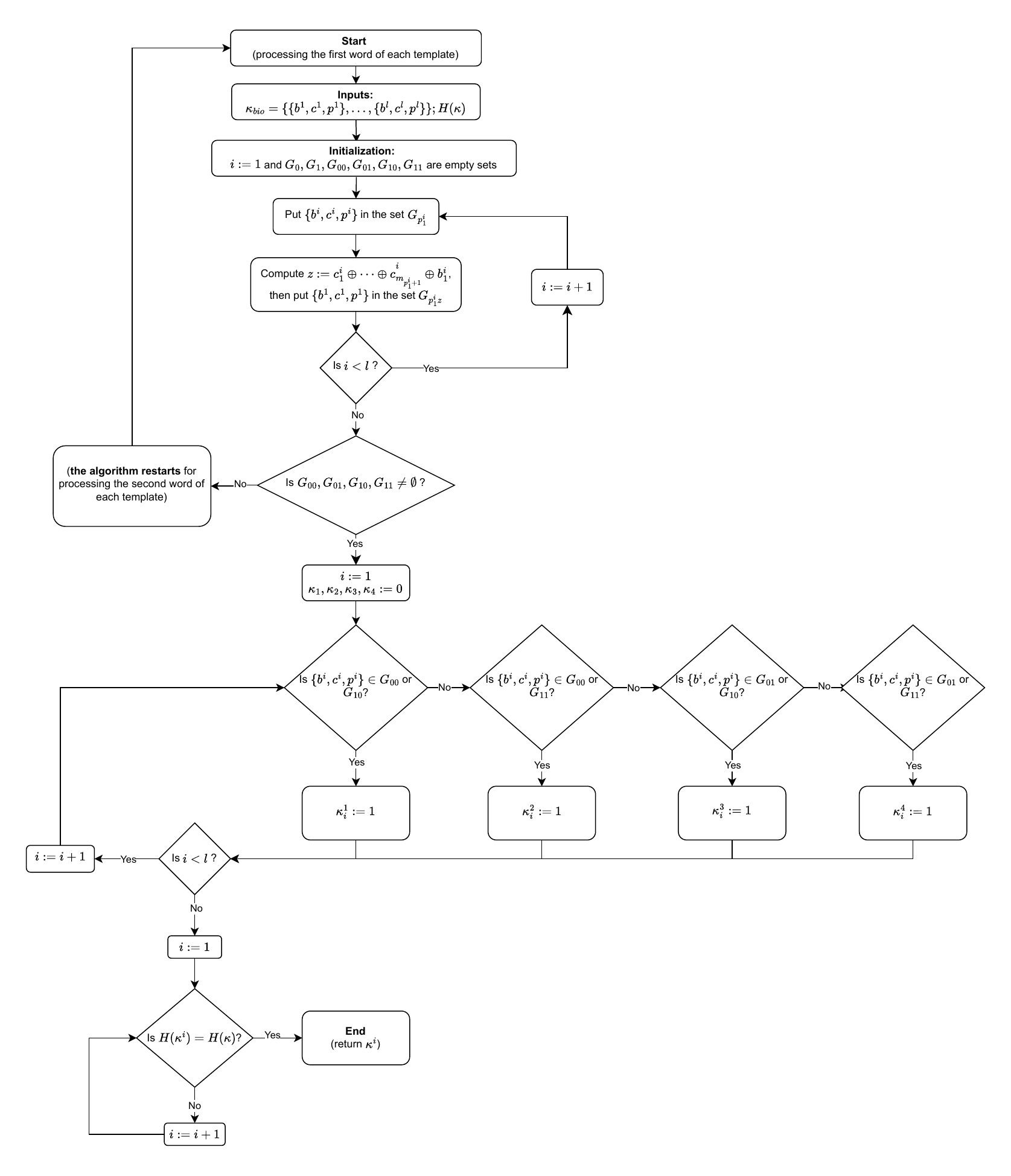}
\end{center}
\caption{Flowchart for retreiving the key based on the first template word.}
\label{fig:flowchart-cryptanalysis}
\end{figure*}

Then, the attacker computes in the first group, the parity of the first 
word of size $m_1$ of all original templates, which is 
$z = c_1^i\oplus\ldots\oplus c_{m_1}^i\oplus b_1^i$,
and puts the corresponding
biocodes into a set $G_{00}$ or $G_{01}$, depending on this parity.
The attacker realizes the same strategy with $G_1$ and obtains
two other sets $G_{10}$ and $G_{11}$. 
At this step, there are two
possibilities, depending on the parity of the first word :
\begin{enumerate}
\item If $G_{00}$, $G_{01}$, $G_{10}$ and $G_{11}$ are non-empty, the attacker 
does not know if the biocodes of these sets have been generated from the genuine 
template or from the fake template, but he knows that all the biocodes of each 
set have been generated from the same template, which is sufficient for the attack. If $G_{00}$ corresponds to the genuine template then $G_{01}$ corresponds to the fake template (and \textit{vice versa}). If $G_{10}$ corresponds to a genuine template then $G_{11}$ corresponds to the fake template (and \textit{vice versa}). As a consequence, the attacker has only $4$ possibilities for the key 
$\kappa$, each of which can be verified against the hash digest $H(\kappa)$. The flowchart provided in Figure~\ref{fig:flowchart-cryptanalysis} offers a visual representation of the algorithm.
\item If $G_{00}$ or $G_{01}$ (resp $G_{10}$ or $G_{11}$) is empty, that means 
that the genuine template and the fake template have the same parity in the 
first $m_1$-bit (resp $m_2$-bit) word. In this case, the attacker cannot 
directly conclude and processes the same strategy for $G_0$ (resp $G_1$) 
with the second word.
\end{enumerate}

In the previous algorithm, we only consider, at the end, the case 
$G_{00}$, $G_{01}$, $G_{10}$, $G_{11}\neq\emptyset$ for simplicity.
If $G_{00}$ or $G_{01}$ is empty, we consider the parity of the second 
word, and separate $G_0$ in two sets $G_{00}$ and $G_{01}$, depending 
on the value of $p_i^2$, then separates these two sets in two
other sets  $G_{000}$,  $G_{001}$,  $G_{010}$,  $G_{011}$,
depending on the value of $c_1^i\oplus\ldots\oplus c_{m_1}^i\oplus b_2^i$,
as previously. A similar strategy is used if $G_{10}$ or $G_{00}$ is empty.

\begin{algorithm}
\caption{\textbf{Cryptanalysis of the CBV scheme}}
\KwIn{$\kappa_{bio}$ =  $\{\{b^1, c^1, p^1\},\ldots ,\{b^l, c^l, p^l\}\}$ and $H(\kappa)$}
\KwOut{the corresponding key $\kappa$}
$G_0$, $G_1$, $G_{00}$, $G_{01}$, $G_{10}$, $G_{11}\leftarrow\emptyset$\\
\For{$i=1$ \KwTo $l$}{
        $j\leftarrow p_1^i$\\
        $G_j\leftarrow G_j\cup \{b^i, c^i, p^i\}$ 
}
\For{$i=1$ \KwTo $l$}{   
        $j\leftarrow p_1^i$\\
        $z\leftarrow c_1^i\oplus\ldots\oplus c_{m_{j+1}}^i\oplus b_1^i$ \\
        $G_{jz}\leftarrow G_{jz}\cup \{b^i, c^i, p^i\}$
}        
\If{$G_{00}$, $G_{01}$, $G_{10}$, $G_{11}\neq\emptyset$}
{
$\kappa^1,\kappa^2,\kappa^3,\kappa^4\leftarrow 0$\\
\For{$i=1$ \KwTo $l$}
{if $\{b^i, c^i, p^i\}\in G_{00}$ or $G_{10}$, $\kappa_i^1\leftarrow 1$ \\
if $\{b^i, c^i, p^i\}\in G_{00}$ or $G_{11}$, $\kappa_i^2\leftarrow 1$ \\
if $\{b^i, c^i, p^i\}\in G_{01}$ or $G_{10}$, $\kappa_i^3\leftarrow 1$ \\
if $\{b^i, c^i, p^i\}\in G_{01}$ or $G_{11}$, $\kappa_i^4\leftarrow 1$  
}
\For{$i=1$ \KwTo $4$}
{If $H(\kappa^i)$ = $H(\kappa)$ \Return $\kappa^i$}
}
\end{algorithm}

\subsection{Numerical examples}

\noindent {\bf Numerical example 1 : }
We consider $n~=~l~=~8$. We suppose the key $\kappa $ = $0x4b$ = $0b1001011$
= $[1,1,0,1,0,0,1,0]$ in little endian notation, $x^e$ = $0xa9$ 
 = $0b10101001$ = $[1,0,0,1,0,1,0,1]$ and $x^f$ = $0xb3$ = $0b10110011$
 = $[1, 1, 0, 0, 1, 1, 0, 1]$. We remark that the parity of the first
 $3$-bit word of $x^e$ and $x^f$ is different, as for the first
 $5$-bits word. Thus, we will have $G_{00}$, $G_{01}$, $G_{10}$, $G_{11}\neq\emptyset$, 
 and the attack will be successful, as detailed below. 
 
\noindent Applying the CBV scheme, we obtain the set of biocodes $\kappa_{bio}$
= $\{B_1,\ldots ,B_8\}$, where\\
$B_1$ = $\{[1,1], [1,1,0,1,0,0,1,0], [3,5]\}$,\\
$B_2$ = $\{[1,0], [1,1,0,0,1,1,1,0], [3,5]\}$,\\
$B_3$ = $\{[1,0], [1,0,1,0,0,0,0,0], [5,3]\}$,\\
$B_4$ = $\{[0,1], [1,1,1,1,1,0,0,0], [3,5]\}$,\\
$B_5$ = $\{[1,0], [1,0,1,1,1,0,0,0], [5,3]\}$,\\
$B_6$ = $\{[0,0], [0,1,1,1,0,0,0,0], [3,5]\}$,\\
$B_7$ = $\{[1,1], [0,0,1,1,1,0,1,0], [5,3]\}$,\\
$B_8$ = $\{[0,1], [0,0,1,0,0,0,1,0], [5,3]\}$.\\

\noindent Remark : for readability purposes, the last component of the biocode is 
not $p$, but the length of each word. We note that $[5,5]$ is not possible
and $[3,3]$ is not present due to our implementation of the  cancelable
transform (the length of the last word is the optimal length).

\noindent From $\kappa_{bio}$ and $H(\kappa)$ we recover $\kappa$ with the
previous algorithm:
we have $G_0$ = $\{B_1, B_2, B_4, B_6\}$ and
$G_1$ = $\{B_3, B_5, B_7, B_8\}$.
Computing the parity bit $z_i$ = $c_1^i\oplus\ldots\oplus c_{m}^i\oplus b_1^i$
for each biocode, we obtain $G_{00}$ = $\{B_6\}$, 
$G_{01}$ = $\{B_1, B_2, B_4\}$, $G_{10}$ = $\{B_7\}$ and $G_{11}$ = $\{B_3, B_5, B_8\}$.\\
These four non empty sets provide four candidate keys, according to notation of 
the previous algorithm, which are :
$\kappa^1$ = $[0,0,0,0,0,1,1,0]$, 
$\kappa^2$ = $[0,0,1,0,1,1,0,1]$,
$\kappa^3$ = $[1,1,0,1,0,0,1,0]$,
and 
$\kappa^4$ = $[1,1,1,1,1,0,0,1]$. The good one is $\kappa^3$, using the
value $H(\kappa)$.\\

\noindent {\bf Numerical example 2 : }
We consider $n = l = 8$. 
We suppose the key $\kappa $ = $0x23$ = $0b100011$
= $[1,1,0,0,0,1,0,0]$ in little endian notation, $x^e$ = $0x34$ 
 = $0b110100$ = $[0,0,1,0,1,1,0,0]$ and $x^f$ = $0xa6$ = $0b10100110$
 = $[0,1,1,0,0,1,0,1]$. We remark that the parity of the first
 $3$-bits word of $x^e$ and $x^f$ is different, but is the same for the first
 $5$-bits word. Thus the attack will be not directly successful, as detailed below. 
 
\noindent Applying the CBV scheme, we obtain the set of biocodes $\kappa_{bio}$
= $\{B_1,\ldots ,B_8\}$, where\\
$B_1$ = $\{[0,1], [0,1,0,0,1,0,1,1], [3,5]\}$,\\
$B_2$ = $\{[1,0], [0,0,0,1,0,1,1,1], [3,5]\}$,\\
$B_3$ = $\{[0,0], [0,0,0,0,0,1,1,0], [5,3]\}$,\\
$B_4$ = $\{[0,0], [0,1,1,1,1,0,0,0], [3,5]\}$,\\
$B_5$ = $\{[1,1], [0,0,0,0,1,1,0,0], [5,3]\}$,\\
$B_6$ = $\{[1,1], [1,1,0,1,0,1,0,1], [5,3]\}$,\\
$B_7$ = $\{[1,1], [0,1,0,0,0,1,0,0], [5,3]\}$,\\
$B_8$ = $\{[0,1], [0,0,1,0,1,0,1,0], [5,3]\}$.\\
 
\noindent From $\kappa_{bio}$ and $H(\kappa)$ we recover $\kappa$ as following.
we have $G_0$ = $\{B_1, B_2, B_4\}$ and 
$G_1$ = $\{B_3, B_5, B_6, B_7, B_8\}$.
Computing the parity bit $z_i$ = $c_1^i\oplus\ldots\oplus c_{m_1}^i\oplus b_1^i$
for each biocode, we obtain $G_{00}$ = $\{B_4\}$, $G_{01}$ = $\{B_1, B_2\}$,
$G_{10}$ = $\{B_3, B_5, B_6, B_7, B_8\}$ and $G_{11} = \emptyset$.\\
Thus, we only consider $G_1$ and we compute the parity bits
$c_{m_2}^i\oplus\ldots\oplus c_{m_1+m_2}^i\oplus b_2^i$ for all biocodes in
$G_1$. We obtain $G_{10}$ = $\{B_3, B_5, B_7, B_8\}$ and $G_{11}$ = $\{B_6\}$.
The four candidate keys, according to notation of the previous algorithm, are :
$\kappa^1$ = $[0,0,1,1,1,0,1,1]$, 
$\kappa^2$ = $[0,0,0,1,0,1,0,0]$,
$\kappa^3$ = $[1,1,1,0,1,0,1,1]$,
and 
$\kappa^4$ = $[1,1,0,0,0,1,0,0]$.
The good one is $\kappa^4$, using the value $H(\kappa)$.\\
 
As a conclusion, the attack cannot be presented on a key of length $l$ = $128$, 
because the length of the corresponding biocode is too large. Nevertheless, the 
attack is very efficient 
and the success rate is 100\% due to the knowledge of $H(\kappa)$.


\section{Discussion}

In this section, we realize a more in-depth analysis on the security of the 
scheme, including a potential strategies to enhance the CBV scheme. 

The performance of the scheme is influenced by the length $n$ of templates, 
the length $l$ of the key, and by the threshold $\tau$. Given a biometric key $k_{bio}$ 
and a fresh template from the same people, the key release algorithm performs $l$
distance computation and fails if only one distance is greater than the threshold $\tau$. 
Furthermore the length $n$ of template is linked to the performance of the
cancelable transform, but in a more complex way, because it is also linked to the
parameter $m$.

In the context of the previous attack, the length of the key $\kappa$ is not linked to
the success rate of the attack because each bit of the key $\kappa_i$ is recovered
independently from the corresponding biocode $\{b^i, c^i, p^i\}$.


The previous attack is based on the assumption that the fake template is the same
for each bits of the key. Consequently, we consider in this section a variant 
where a distinct fake template is used for generating a biocode for each 
bit of the key $\kappa$, during the key-binding procedure.

In a first time, the attacker separates the $l$ biocodes into two groups 
$G_0$ and $G_1$, depending on the value of the bit $p_1^i$, as previously.
Without loss of generalities, we suppose that $l/2$ biocodes are generated from 
the genuine template and $l/2$ biocodes are generated from the distinct fake 
templates, modelized by a random template.
The attacker computes also the parity of the first word for each set
$G_0$ and $G_1$. In this scenario, there will be
around $3/4$ with the same parity and $1/4$ with an other parity
(all biocodes generated from the genuine template and around half
of biocodes generated from a fake template). 
Thus, the attacker knows half of biocodes generated from a fake template. 

In a second time, the attacker removes the $l/4$ identified biocodes of $G_0$ 
and $G_1$ and separates each group $G_0$ and $G_1$ into 
two groups $G_{00}, G_{01}, G_{10}, G_{11}$, depending on the value of the bit 
$p_2^i$, for $i$ between $1$ and $l$ (except for the $l/4$ identified biocodes).
Then the attacker computes the parity of the second word on each group and 
identifies the biocodes from a fake template as previously.
 There will be around $5/6$ with the same parity and $1/6$ with an other parity.
The same strategy is processed until all the biocodes from a fake template 
are identified.
Nevertheless, a brute force attack should be realized at the end of this second 
attack, due to the impossibility to realize statistics on a small amount of data. 

\noindent {\bf Numerical example with $l$ = $128$ : } It is not possible
to do statistical tests on small amount of values as in the previous section 
with $l$ = $n$ = $8$. Remember that $l$ is the length of the key, 
so $l\geq 128$ for any use in cryptography.
Without loss of generality, we only consider the average case in this example :

At the first step, for the two 
groups, each of size $64$, we have $16$ biocodes with a given 
parity for the first word and $48$ with the other parity, so we 
identify $32$ biocodes from the fake template.
In the second step, for the four groups, each of size $24$, we have 
$4$ biocodes with a given parity for the first word 
and $22$ with the other parity, so we identify $16$ biocodes.
In the third step, for the eight groups, each of size $10$, we have 
$1$ biocode with a given parity for the first word 
and $9$ with the other parity, so we identify $8$ biocodes.
At this step of the algorithm, we have already 56 biocodes 
from the fake template over 64. The attacker recovers the key from
$H(\kappa)$ with a brute force
attack on $\kappa$ in $\binom{128}{8}\simeq 2^{40}$ tests.

The previous example uses a similar assumption to the linkability 
attack. Nevertheless, this assumption is not related to the 
FMR/FNMR of the biometric database, but only to the difference
between the biocode generated from the original template and the
biocodes generated from fake templates during the key-binding
algorithm. If these biocodes are
close to the genuine biocode, the key space of the attack is higher,
but, in the same time, the success probability of the key-release
algorithm decreases for a genuine person.

\begin{algorithm}
\caption{\textbf{Cryptanalysis of the modified CBV scheme}}
\KwIn{$\kappa_{bio}$ =  $\{\{b^1, c^1, p^1\},\ldots ,\{b^l, c^l, p^l\}\}$ and 
$H = H(\kappa)$}
\KwOut{the corresponding key $\kappa$}
$G_0$, $G_1$, $G_{00}$, $G_{01}$, $G_{10}$, $G_{11}\leftarrow\emptyset$\\
\For{$i=1$ \KwTo $l$}{
        $j\leftarrow p_i^1$\\
        $G_j\leftarrow G_j\cup \{b^i, c^i, p^i\}$ 
}
\For{$i=1$ \KwTo $l$}{   
        $j\leftarrow p_i^1$\\
        $z\leftarrow c_1^i\oplus\ldots\oplus c_{m}^i\oplus b_1^i$\\
        $G_{jz}
        \leftarrow G_{jz}\cup \{b^i, c^i, p^i\}$
}  
$\kappa \leftarrow 2^l-1$\\

\ForAll{$\{b^i, c^i, p^i\}\in \min(G_{00}, G_{01})$}
{$\kappa_i\leftarrow 0$}
\ForAll{$\{b^i, c^i, p^i\}\in \min(G_{10}, G_{11})$}{$\kappa_i\leftarrow 0$}
Comeback to step 1 to recover the unknown bits of $\kappa$ or use an
exhaustive search.\\
If $H(\kappa)= H$ \Return $\kappa$
\end{algorithm}

\section{Conclusion}
\label{sec:conclusion}

This paper proposed an efficient cryptanalysis of a recent key-binding scheme, 
by identifying a vulnerability of the  underlying cancelable transformation.
The key is recovered from only one biocode without additional assumptions
on the knowledge of the attacker.
A modified version of this scheme is also analyzed, which may appear more secure at a 
glance, but upon closer examination fails to offer enhancements in security.

The effectiveness and practical impact of the proposed cryptanalysis are demonstrated through numerical examples because the attacks directly recover the cryptographic key from the binary biocode, regardless of the original biometric data. The cryptanalysis retrieves a 128-bit key generated from a random biometric data in less than 1 second using Python on both the original
scheme and the variant.

The main drawbacks of this scheme are the vulnerability of the cancelable transform
and the absence of any security proof, while alternative solutions exist. Nevertheless 
most of these solutions have not 
been implemented on real biometric data and also require a more in-depth
analysis. A such efficient and secure key-binding scheme is crucial, 
considering the biometric data protection issues.

\section*{Acknowledgement}
The authors acknowledge the support of the French Agence Nationale de la Recherche (ANR), under grant ANR-20-CE39-0005 (project PRIVABIO).


\bibliographystyle{elsarticle-num} 
\bibliography{fuzzy}

\begin{thebibliography}{10}
\expandafter\ifx\csname url\endcsname\relax
  \def\url#1{\texttt{#1}}\fi
\expandafter\ifx\csname urlprefix\endcsname\relax\def\urlprefix{URL }\fi
\expandafter\ifx\csname href\endcsname\relax
  \def\href#1#2{#2} \def\path#1{#1}\fi

\bibitem{jnn08}
A.~K. Jain, K.~Nandakumar, A.~Nagar, Biometric template security, EURASIP
  Journal on Advances in Signal Processing (2008) 1--17.

\bibitem{ru11}
C.~Rathgeb, A.~Uhl, A survey on biometric cryptosystems and cancelable
  biometrics, EURASIP Journal on Information Security (2011).

\bibitem{mrtvb22}
P.~Melzi, C.~Rathgeb, R.~Tolosana, R.~Vera-Rodriguez, C.~Busch, An overview of
  privacy-enhancing technologies in biometric recognition, coRR abs/2206.10465
  (2022).

\bibitem{onr20}
O.~Ouda, K.~Nandakumar, A.~Ross, Cancelable biometrics vault: A secure
  key-binding biometric cryptosystem based on chaffing and winnowing, in:
  International Conference on Pattern Recognition (ICPR), 2020.

\bibitem{rcb01}
N.~K. Ratha, J.~H. Connell, R.~M. Bolle, Enhancing security and privacy in
  biometrics-based authentication system, IBM Systems J. 37~(11) (2001)
  2245--2255.

\bibitem{tng04}
A.~B.~J. Teoh, D.~N.~C. Ling, A.~Goh, Biohashing: Two factor authentication
  featuring fingerprint data and tokenised random number, Pattern Recognition
  37~(11) (2004) 2245--2255.

\bibitem{tgn06}
A.~B.~J. Teoh, A.~Goh, D.~N.~C. Ling, Random multispace quantization as an
  analytic mechanism for biohashing of biometric and random identity inputs,
  IEEE Transactions on Pattern Analysis and Machine Intelligence 28~(12) (2006)
  1892--1901.

\bibitem{tkl08}
A.~B.~J. Teoh, Y.~W. Kuan, S.~Lee, Cancellable biometrics and annotations on
  biohash, Pattern Recognition 41~(6) (2008) 2034--2044.

\bibitem{fyj10}
Y.~C. Feng, P.~C. Yuen, A.~K. Jain, A hybrid approach for generating secure and
  discriminating face template, IEEE Transactions on Information Forensics and
  Security 5~(1) (2010) 103--117.

\bibitem{ppcr11}
J.~K. Pillai, V.~M. Patel, R.~Chellappa, N.~K. Ratha, Secure and robust iris
  recognition using random projections and sparse representations, IEEE
  Transactions on Pattern Analysis Machine Intelligence 33~(9) (2011)
  1877--1893.

\bibitem{wp10}
Y.~Wang, K.~N. Plataniotis, An analysis of random projection for changeable and
  privacy-preserving biometric verification, IEEE Transations on Systems, Man,
  and Cybernetics, Part B 40~(5) (2010) 1280--1293.

\bibitem{jl84}
W.~Johnson, J.~Lindenstrauss, Extensions of lipschitz maps into a hilbert
  space, in: Contemporary Mathematics, 1984, pp. 189--206.

\bibitem{nnj10}
A.~Nagar, K.~Nandakumar, A.~K. Jain, Biometric template transformation: A
  security analysis, in: SPIE, Electronic Imaging, Media Forensics and Security
  XII, 2010.

\bibitem{fly14}
Y.~C. Feng, M.-H. Lim, P.~C. Yuen, Masquerade attack on transform-based
  binary-template protection based on perceptron learning, Pattern Recognition
  47~(9) (2014) 3019--3033.

\bibitem{tkae16}
B.~Topcu, C.~Karabat, M.~Azadmanesh, H.~Erdogan, Practical security and privacy
  attacks against biometric hashing using sparse recovery, EURASIP Journal on
  Advances in Signal Processing 2016~(1) (2016) 100.

\bibitem{lcr13}
P.~Lacharme, E.~Cherrier, C.~Rosenberger, Reconstruction attack on biohashing,
  in: International Conference on Security and Cryptography (SECRYPT),
  SciTePress, 2013, pp. 363--370.

\bibitem{grgfo13}
J.~Galbally, A.~Ross, M.~Gomez-Barrero, J.~Fierrez, J.~Ortega-Garcia, Iris
  image reconstruction from binary templates: An efficient probabilistic
  approach based on genetic algorithms, Computer Vision and Image Understanding
  117~(10) (2013) 1512--1525.

\bibitem{rgb15}
A.~Rozsa, A.~E. Glock, T.~E. Boult, Genetic algorithm attack on minutiae-based
  fingerprint authentication and protected template fingerprint systems, in:
  IEEE Computer Vision and Pattern Recognition Workshops (CVPR), 2015.

\bibitem{djt19}
X.~Dong, Z.~Jin, A.~B.~J. Teoh, A genetic algorithm enabled similarity-based
  attack on cancellable biometrics, in: IEEE International Conference on
  Biometrics: Theory, Applications and Systems (BTAS), 2019, pp. 1--8.

\bibitem{rbbb14}
C.~Rathgeb, F.~Breitinger, C.~Busch, H.~Baier, On application of bloom filters
  to iris biometrics, IET Biometrics 3~(4) (2013) 207--218.

\bibitem{rgrfb14}
M.~Gomez-Barrero, C.~Rathgeb, J.~Galbally, J.~Fierrez, C.~Busch, Protected
  facial biometric templates based on local gabor patterns and adaptive bloom
  filters, in: International Conference on Pattern Recognition, 2014, pp.
  4483--4488.

\bibitem{bmr15}
J.~Bringer, C.~Morel, C.~Rathgeb, Security analysis of bloom filter-based iris
  biometric template protection, in: International Conference on Biometrics
  (ICB), 2015, pp. 527--534.

\bibitem{rccb07}
N.~K. Ratha, S.~Chikkerur, J.~H. Connell, R.~M. Bolle, Generating cancelable
  fingerprint templates, IEEE Transactions on Pattern Analysis and Machine
  Intelligence 29 (2007) 561–572.

\bibitem{fscz08}
Q.~Feng, F.~Su, A.~Cai, F.~Zhao, Cracking cancelable fingerprint template of
  ratha, in: International Symposium on Computer Science and Computational.
  Technology, 2008, pp. 572--575.

\bibitem{c02}
M.~Charikar, Similarity estimation techniques from rounding algorithms, in: ACM
  Symposium on Theory of Computing, 2002, pp. 380--388.

\bibitem{jhlkt18}
Z.~Jin, J.-Y. Hwang, Y.-L. Lai, S.~Kim, A.~B.~J. Teoh, Ranking-based locality
  sensitive hashing-enabled cancelable biometrics: Index-of-max hashing, {IEEE}
  Transactions on Information Forensics and Security (TIFS) 13~(2) (2018)
  393--407.

\bibitem{gkla20}
L.~Ghammam, K.~Karabina, P.~Lacharme, K.~Atighehchi, A cryptanalysis of two
  cancelable biometric schemes based on index-of-max hashing, IEEE Transactions
  on Information Forensics and Security 15 (2020) 2869--2880.

\bibitem{drs04}
Y.~Dodis, L.~Reyzin, A.~D. Smith, Fuzzy extractors: How to generate strong keys
  from biometrics and other noisy data, in: Advances in Cryptology (Eurocrypt),
  Vol. 3027 of LNCS, Springer, 2004, pp. 523--540.

\bibitem{jw99}
A.~Juels, M.~Wattenberg, A fuzzy commitment scheme, in: ACM Conference on
  Computer and Communications Security (CCS), 1999, pp. 28--36.

\bibitem{js02}
A.~Juels, M.~Sudan, A fuzzy vault scheme, in: IEEE International Symposium on
  Information Theory (ISIT), 2002, p. 408.

\bibitem{had06}
F.~Hao, R.~Anderson, J.~Daugman, Combining crypto with biometrics effectively,
  IEEE Transactions on information forensics and security (TIFS) 55~(9) (2006).

\bibitem{yv07}
S.~Yang, I.~Verbauwhede, Secure iris verification, in: IEEE International
  Conference on Acoustics, Speech, and Signal Processing (ICASSP), 2007, pp.
  133--136.

\bibitem{bcckz08}
J.~Bringer, H.~Chabanne, G.~D. Cohen, B.~Kindarji, G.~Zémor, Theoretical and
  practical boundaries of binary secure sketches, IEEE Transactions on
  Information Forensics and Security (TIFS) 3~(4) (2008).

\bibitem{njp07}
K.~Nandakumar, A.~K. Jain, S.~C. Pankanti, Fingerprint-based fuzzy vault :
  Implementation and performance, IEEE Transactions on Information Forensics
  and Security (TIFS) 2~(4) (2007).

\bibitem{iw10}
A.~Ignatenko, F.~M.~J. Willems, Information leakage in fuzzy commitment
  schemes, IEEE Transactions on information forensics and security (TIFS) 5~(2)
  (2010).

\bibitem{zkvb11}
X.~Zhou, A.~Kuijper, R.~Veldhuis, C.~Busch, Quantifying privacy and security of
  biometric fuzzy commitment, in: IEEE International Joint Conference on
  Biometrics (IJCB), 2011, pp. 1--8.

\bibitem{b04}
X.~Boyen, Reusable cryptographic fuzzy extractors, in: ACM Conference on
  Computer and Communications Security (CCS), 2004, pp. 82--91.

\bibitem{stp09}
K.~Simoens, P.~Tuyls, B.~Preneel, Privacy weaknesses in biometric sketches, in:
  IEEE Symposium on Security and Privacy, 2009, pp. 188--203.

\bibitem{kbkbv11}
E.~J.~C. Kelkboom, J.~Breebaart, T.~A.~M. Kevenaar, I.~Buhan, R.~N.~J.
  Veldhuis, Preventing the decodability attack based cross-matching in a fuzzy
  commitment scheme, IEEE Transactions on Information Forensics and Security
  (TIFS) 6~(1) (2011) 107--121.

\bibitem{t14}
B.~Tams, Decodability attack against the fuzzy commitment scheme with public
  feature transforms, coRR abs/1406.1154 (2014).

\bibitem{ba13}
M.~Blanton, M.~Aliasgari, Analysis of reusability of secure sketches and fuzzy
  extractors, IEEE Transactions on Information Forensics and Security 8~(9)
  (2013) 1433--1445.

\bibitem{tmm15}
B.~Tams, P.~Mihailescu, A.~Munk, Security considerations in minutiae-based
  fuzzy vaults, IEEE Transactions on Information Forensics and Security 10~(5)
  (2015) 985--998.

\bibitem{rtwb16}
C.~Rathgeb, B.~Tams, J.~Wagner, C.~Busch, Unlinkable improved multi-biometric
  iris fuzzy vault, EURASIP Journal on Information Security (2016) 26.

\bibitem{fmr13}
B.~Fuller, X.~Meng, L.~Reyzin, Computational fuzzy extractors, in: Advances in
  Cryptology (Asiacrypt), Vol. 8269 of LNCS, Springer, 2013, pp. 174--193.

\bibitem{acek17}
D.~Apon, C.~Cho, K.~Eldefrawy, J.~Katz, Efficient, reusable fuzzy extractors
  from lwe, in: International Conference on Cyber Security Cryptography and
  Machine Learning (CSCML), 2017, pp. 1--18.

\bibitem{wl18}
Y.~Wen, S.~Liu, Robustly reusable fuzzy extractor from standard assumptions,
  in: Advances in Cryptology (ASIACRYPT), Vol. 11274 of LNCS, Springer, 2018,
  pp. 459--489.

\bibitem{abccfgs18}
Q.~Alam\'elou, P.-E. Berthier, C.~Cachet, S.~Cauchie, B.~Fuller, P.~Gaborit,
  S.~Simhadri, Pseudoentropic isometries: A new framework for fuzzy extractor
  reusability, in: ACM Asia Conference on Computer and Communications Security
  (AsiaCCS), 2018, pp. 673--684.

\bibitem{cfprs21}
R.~Canetti, B.~Fuller, O.~Paneth, L.~Reyzin, A.~D. Smith, Reusable fuzzy
  extractors for low-entropy distribution, Journal of Cryptology 34~(1) (2021).

\bibitem{ssf19}
S.~Simhadri, J.~Steel, B.~Fuller, Cryptographic authentication from the iris,
  in: International Conference on Information Security, 2019, pp. 465--485.

\bibitem{uyckl21}
E.~Uzun, C.~Yagemann, S.~Chung, V.~Kolesnikov, W.~Lee, Cryptographic key
  derivation from biometric inferences for remote authentication, in: ACM Asia
  Conference on Computer and Communications Security (AsiaCCS), 2021, pp.
  629--643.

\bibitem{wald1944cumulative}
A.~Wald, On cumulative sums of random variables, The Annals of Mathematical
  Statistics 15~(3) (1944) 283--296.

\end{thebibliography}


\end{document}